\documentclass[aps,prb,final,twocolumn,superscriptaddress]{revtex4-1}

\usepackage{graphicx}
\usepackage{epsfig}
\usepackage{natbib}

\newcommand{\CrNbS}{$\mathrm{CrNb}_3\mathrm{S}_6$}

\begin{document}


\title{Understanding the $H-T$ phase diagram of the monoaxial helimagnet}


\author{Victor Laliena}
\email[]{laliena@unizar.es}
\affiliation{Instituto de Ciencia de Materiales de Arag\'on 
(CSIC -- University of Zaragoza), 
C/Pedro Cerbuna 12, 50009 Zaragoza, Spain}

\author{Javier Campo}
\email[]{javier.campo@csic.es}
\affiliation{Instituto de Ciencia de Materiales de Arag\'on 
(CSIC -- University of Zaragoza), 
C/Pedro Cerbuna 12, 50009 Zaragoza, Spain}
\affiliation{Centre for Chiral Science, Hiroshima University, Higashi-Hiroshima, 
Hiroshima 739-8526, Japan}

\author{Yusuke Kousaka}
\affiliation{Department of Chemistry, Faculty of Science, Hiroshima University, 
Higashi-Hiroshima, Hiroshima 739-8526, Japan}
\affiliation{Centre for Chiral Science, Hiroshima University, Higashi-Hiroshima, 
Hiroshima 739-8526, Japan}


\date{August 12, 2016}

\begin{abstract}
Some unexpected features of the phase diagram of the monoaxial helimagnet in presence of
an applied magnetic field perpendicular to the chiral axis are theoretically predicted.
A rather general hamiltonian with long range Heisenberg exchange and Dzyaloshinskii--Moriya 
interactions is considered. The continuum limit simplifies the free energy, 
which contains only a few parameters which in principle are determined 
by the many parameters of the hamiltonian, although in practice they may be tuned to
fit the experiments. The phase diagram  
contains a Chiral Soliton Lattice phase and a forced ferromagnetic phase separated by a line 
of phase transitions, which are of second order at low $T$ and of first order in the vicinity of the 
zero-field ordering temperature, and are separated by a tricritical point.
A highly non linear Chiral Soliton Lattice, in which many harmonics contribute appreciably to the
spatial modulation of the local magnetic moment, develops only below the tricritical temperature,
and in this case the scaling shows a logarithmic behaviour
similar to that at $T=0$, which is a universal feature of the Chiral Soliton Lattice. 
Below the tricritical temperature,
the normalized soliton density curves are found to be independent of $T$, in agreement with 
the experimental results of magnetorresistance curves, while above the tricritical temperature
they show a noticeable temperature dependence.
The implications in the interpretation of experimental results of \CrNbS\ are discussed.
\end{abstract}

\pacs{111222-k}
\keywords{Helimagnet, Dzyaloshinskii--Moriya interaction, Chiral Soliton Lattice}

\maketitle


\section{Introduction}

Chiral magnets are very promising ingredients for spintronic based devices since they support 
peculiar magnetic textures that affect the charge and spin transport properties in different ways. 
As these magnetic textures can be deeply altered by magnetic fields,
the transport properties can be magnetically controlled \cite{Wolf01,Zutic04}. 
The chiral topological nature of these magnetic textures endows them with a protective mechanism, 
since they cannot be continuously deformed to more conventional magnetic states like ferromagnetic 
order \cite{Fert13}.  
This robustness makes chiral magnets excelent candidates as the main components of information storage 
devices \cite{Romming13}. 
On the other hand, it is worthwile to stress that, besides the applications to spintronics, chiral magnets are 
interesting from a fundamental point of view, as chiral symmetry and its breaking and restoration 
are ubiquous phenomena appearing virtually in any domain of science, from particle physics to astrophysics, 
and including chemistry, biology, and geology \cite{Wagniere07}.

In the monoaxial helimagnet, the competition between the ferromagnetic (FM) and Dzyaloshinskii--Moriya (DM) 
interactions at low $T$ results in a magnetic helix propagating with period $L_0$ along a crystallographic 
axis, which is called here the DM axis. At a certain ordering temperature, $T_0$,
a magnetic transition to a paramagnetic (PM) phase takes place.
For temperatures lower than $T_0$, application of a magnetic field perpendicular to the DM axis 
deforms the helix and a Chiral Soliton Lattice (CSL) appears
\cite{Kishine15,Dzyal58,Dzyal64,Izyumov84,Kishine05}. This CSL, which is realised \cite{Togawa12} in 
\CrNbS, supports dynamical modes like coherent sliding \cite{Kishine12} and gives 
rise to phenomena very interesting for spintronics, like spin motive forces \cite{Ovchinnikov13} 
and tuneable magnetoresistence \cite{Kishine11,Togawa13,Ghimire13,Bornstein15}. 
By increasing the field the period of the CSL increases and, eventually, as the period diverges, 
a transition takes place continuously to a forced FM state (FFM).
The nature of the transition in the vicinity of $T_0$ is not fully understood and considerable effort 
is being devoted to clarify this interesting question
\cite{Tsuruta16,Bornstein15,Togawa13,Ghimire13,Chapman14,Shinozaki15}.

DeGennes \cite{DeGennes75} introduced a classification of the continuous transitions that take
place between spatially homogeneous and  modulated states. He named \textit{nucleation transitions} 
those in which the period of the modulated state diverges when the transition point is approached
from the modulated phase. Transitions in which the intensity of the Fourier modes with non-zero
wave-vector tend to zero while the fundamental wave-vector remains non-zero were called 
\textit{instability transitions}. 
The transition mechanisms for nucleation and instability transitions are 
very different. In the monoaxial helimagnet, the transition between the CSL and the FFM states  
as a perpendicular magnetic field increases at zero temperature
is of nucleation type \cite{Dzyal64}. On the other hand, mean field theory predicts an instability type 
continuous transition
at the ordering temperature $T_0$ for zero field. Hence, by varying the temperature from $0$
to $T_0$ the transition changes from nucleation to instability type. 
How this change of regime takes place is a very interesting question which may also have
interesting phenomenological consequences.

In this paper the magnetic phase diagram with magnetic field perpendicular to the DM axis
and the nature of the transition from the
CSL to the FFM states are theoretically studied. The question posed in the above paragraph,
how the transition changes from nucleation to instability type, is answered.
The thermal fluctuations are treated classically 
and therefore the results are not valid at very low $T$, where it is well known that a quantum 
treatment of thermal fluctuations is necessary, for instance, to reproduce the behaviour of the 
specific heat. At T=0, however, the semiclassical approximation seems to describe well the ground
state structure of these kind of systems.


\section{Model and method of solution \label{sec:model}}

Let us consider a classical spin system with FM exchange and monoaxial DM interactions, 
and single-ion easy-plane anisotropy,  at temperature $T$ and in presence of an applied magnetic 
field $\vec{H}$. The hamiltonian, $\mathcal{H}$, is the sum of four terms
\begin{eqnarray}
\mathcal{H}_{\mathrm{FM}} &=& 
- \sum_{\vec{r},\vec{r}^{\,\prime}}J_{\vec{r}^{\,\prime}}\vec{S}_{\vec{r}}\cdot\vec{S}_{\vec{r}+\vec{r}^{\,\prime}}, 
\label{eq:hamilFM} \\
\mathcal{H}_{\mathrm{DM}} &=& - \sum_{\vec{r},z^\prime}D_{z^\prime}\hat{z}\cdot(\vec{S}_{\vec{r}}\times 
\vec{S}_{\vec{r}+z^\prime\hat{z}}), 
\label{eq:hamilDM} \\
\mathcal{H}_{\mathrm{A}} &=& K \sum_{\vec{r}}(\hat{z}\cdot\vec{S}_{\vec{r}})^2, \label{eq:hamilAnis} \\
\mathcal{H}_{\mathrm{Z}} &=& -g\mu_\mathrm{B}\vec{H}\cdot\sum_{\vec{r}}\vec{S}_{\vec{r}}, \label{eq:hamilZeeman}
\end{eqnarray}
where $\vec{r}$ runs over the sites of the magnetic ions lattice 
and $\vec{r}^{\,\prime}$ (with $\vec{r}^{\,\prime}\ne 0$) over the differences between the sites of the
magnetic ions lattice. Therefore, $J_{\vec{r}^{\,\prime}}$ is the long range Heisenberg coupling constant between 
the spins at sites $\vec{r}$ and $\vec{r}+\vec{r}^{\,\prime}$. 
We denote by $\hat{z}$ the unit vector pointing along the DM axis and by $z^\prime\hat{z}$ the
vectors that join pairs of ions which interact via the long range DM interaction, 
and thus $D_{z^\prime}$ is the long range DM coupling constant between the magnetic ions
at positions $\vec{r}$ and $\vec{r}+z^\prime\hat{z}$.
Notice that we ignore any temperature dependence of the magnetic coupling constants.
Finally, $K$ is the single-ion anisotropy strength, $g$ the gyromagnetic factor of the magnetic ion, and
$\mu_\mathrm{B}$ the Bohr magneton.
This is a general model with long range interactions that contains
(infinitely) many free parameters. Nevertheless, we will see that in the continuum limit the free energy contains
only a small number of effective parameters which in principle might be computed from the microscopic 
parameters entering the hamiltonian. It is widely accepted that the magnetic properties of 
compounds like \CrNbS\ are described by the kind of hamiltonians proposed here\cite{Kishine15,Chapman14}.
In what follows we use a cartesian coordinate system with 
right-handed axes $(\hat{x},\hat{y},\hat{z})$ and the convenient notation for the coordinates,
$(x,y,z)=(x_1,x_2,x_3)$, and derivatives, $\partial_i = \partial/\partial x_i$.

The statistical properties are given by the partition function, 
\begin{equation}
\mathcal{Z}=\int [d\hat{n}]\exp(-\mathcal{H}[\hat{n}]/k_\mathrm{B}T), 
\end{equation}
where $\hat{n}=\vec{S}/S$ is a unit vector in the spin direction, 
$[d\hat{n}]=\prod_{\vec{r}} d^2\hat{n}_{\vec{r}}$,
and $d^2\hat{n}$ is the invariant measure over the unit sphere. 
To evaluate $\mathcal{Z}$ we use the variational mean field approximation, 
which is the lowest order term of a systematic loop 
expansion \cite{ZinnJustin97}. It has been succesfully applied to
the study of the double-exchange model of itinerant ferromagnetism\cite{Laliena01,laliena01b,Laliena02} and,
in combination with
ab-initio techniques, to the study of the temperature dependence of thermodynamic quantities
in itinerant ferromagnets\cite{Gyorffy85,Staunton86,Staunton06}. Let us describe it 
briefly. Consider the trial ``hamiltonian''
$\mathcal{H}_0=\sum_{\vec{r}}\vec{M}_{\vec{r}}\cdot\hat{n}_{\vec{r}}$, where the mean field 
$\vec{M}_{\vec{r}}$ is, in principle, arbitrary, and define the expectation value
$\langle\cdot\rangle_0$ of any functional $\mathcal{O}$ of $\{\hat{n}_{\vec{r}}\}$ as 
\begin{equation}
\langle\mathcal{O}\rangle_0=(1/\mathcal{Z}_0)\int [d\hat{n}] \exp(-\mathcal{H}_0)\mathcal{O}, 
\end{equation}
where $\mathcal{Z}_0=\int [d\hat{n}]\exp(-\mathcal{H}_0)$. Obviously we have
$\mathcal{Z}=\mathcal{Z}_0 \langle\exp(\mathcal{H}_0-\mathcal{H}/k_\mathrm{B}T)\rangle_0$, and, due to the 
convexity of the exponential function, the Jensen inequality\cite{Jensen1906} holds in the form 
\begin{equation}
\mathcal{Z}\geq\mathcal{Z}_0\exp(\langle\mathcal{H}_0\rangle_0-\langle\mathcal{H}\rangle_0/k_\mathrm{B}T).
\end{equation}
In terms of the free energy, $\mathcal{F}=-k_\mathrm{B}T\ln\mathcal{Z}$, the inequality is sometimes called
the Jensen-Feynman inequality or the Gibbs-Bogoliuvov inequality, and reads 
$\mathcal{F}\leq\mathcal{F}_0$, where 
\begin{equation}
\mathcal{F}_0=\langle\mathcal{H}\rangle_0 - k_\mathrm{B}T(\langle\mathcal{H}_0\rangle_0-\ln\mathcal{Z}_0) 
\end{equation}
is a functional of the mean field configuration. Hence, the best approximation to the true free energy 
$\mathcal{F}$ is obtained with the mean field configuration which minimizes
$\mathcal{F}_0$. The key point is that it is easy to compute $\mathcal{F}_0$, which in the
continuum limit reads 
\begin{equation}
\mathcal{F}_0=(JS^2a^2/v)\int d^3r f_0(\vec{r}), \label{eq:F0contLim}
\end{equation}
where $J=\sum_{\vec{r}^{\,\prime}} J_{\vec{r}^{\,\prime}}$ is an effective Heisenberg interaction constant,  
$v$ is the volume of the elementary cell, and
$a$ an effective average distance between magnetic ions in the $\hat{z}$ direction, in which
the ion distances are weighted by the magnetic
exchange couplings:
\begin{equation}
a^2 = (1/J)\sum_{\vec{r}^{\,\prime}} z^{\prime 2} J_{\vec{r}^{\,\prime}}. \label{eq:a}
\end{equation}
The free energy density $f_0(\vec{r})$ has the form
\begin{widetext}
\begin{equation}
f_0=\frac{1}{2}\sum_i(\xi_i\partial_i\vec{m})^2-\frac{\mu^2q_0^2}{2}m^2
-q_0\hat{z}\cdot(\vec{m}\times\partial_z\vec{m})
+\gamma\left[m/M+(1-3m/M)M_z^2/M^2\right]
-\vec{\beta}\cdot\vec{m}
-\alpha\left[\ln(\sinh M/M) - M m\right], \label{eq:f0}
\end{equation}
\end{widetext}
where 
$\xi_i = (1/Ja^2)\sum_{\vec{r}^{\,\prime}}x_i^{\prime\,2}J_{\vec{r}^{\,\prime}}$,
so that $\xi_z=1$ and by symmetry $\xi_x=\xi_y=\xi$ (see appendix~\ref{sec:contLim}).
The parameter $q_0$, which has the dimension of inverse length, measures the importance of the
DM interaction relative to the exchange interaction, and thus sets the spatial scale of the 
modulation of $\vec{m}$. It is given by  
\begin{equation}
q_0 = (D/Ja^2)\sum_{z^\prime} z^\prime D_{z^\prime}/D, \label{eq:q0}
\end{equation}
where $D=\sum_{z^\prime}D_{z^\prime}$ is the effective DM coupling constant. 
Finally, $\mu^2 = 2/a^2q_0^2$, and the parameters $\gamma$, $\vec{\beta}$, and $\alpha$ are proportional 
to $K$, $\vec{H}$, and $T$, respectively, and are given in Eqs.~(\ref{eq:gamma})-(\ref{eq:alpha})
of the appendix.
We also use the notation $\vec{m}=\langle\hat{n}\rangle_0=F\vec{M}$ and
\begin{equation}
F=\coth(M)/M-1/M^2.
\end{equation}
Some details about the continuum limit and the origin of the parameters appearing in $f_0$ are given in 
appendix~\ref{sec:contLim}.

In the case that only first neighbours interactions are present, calling $J_z$ and $\xi J_z$ the
exchange couplings in the DM direction ($\hat{z}$) and in the transverse directions
($\hat{x}$ and $\hat{y}$), respectively, we get $J=(1+2\xi)J_z$, 
$a^2=a_0^2/(1+2\xi)$, $\mu^2=2(1+2\xi)/(q_0^2 a_0^2)$, and  $q_0=D/J_za_0$,
where $a_0$ is the inter-ion distance in the $\hat{z}$ direction.
These are the expressions for $\mu^2$ and $q_0$ given in a first short version of this paper\footnote{
V. Laliena, J. Campo, Y. Kousaka, and K. Inoue, arXiv:1603.06362v1.}.

It is worthwhile to stress that, in spite of the complexity of the microscopic hamiltonian, the 
continuum limit of the mean field free energy contains only a few independent parameters which
can be tuned to describe the experimental results. The continuum limit is thus universal in the sense 
that it describes the physics of very complicated hamiltonians 
with a large number of parameters in terms of an effective free energy with only a few parameters.
It is a valid approximation when the local magnetic moment 
$\vec{m}_{\vec{r}}$ varies only appreciably over distances long in comparison to 
the range of the magnetic interactions.
This condition clearly implies $q_0a\ll 1$,
since $a$ measures the range of the magnetic interactions and $q_0$ sets the scale of the spatial
variation of $\vec{m}_{\vec{r}}$. The continuum limit is obtained by expanding 
$\vec{m}_{\vec{r}+\vec{r}^{\,\prime}}$ in Taylor series around $\vec{r}$ and keeping only the terms up to
two derivatives. 
The drastic reduction in the number of parameters is due to the fact that in the continuum limit we 
only keep the lowest order derivatives of $\vec{m}_{\vec{r}}$, while the exact free energy contains
terms with any number of derivatives. We also used symmetry to further reduce the number of parameters.
The neglected terms are of the order $q_0^3a^3=(2/\mu^2)^{3/2}$ or higher and the accuracy of the continuum
limit is of the order of $1/\mu^3$ (see Appendix~\ref{sec:contLim}). 
Equation~(\ref{eq:q0}) shows that $q_0a$ scales as the ratio of the 
DM and Heisenberg coupling constants, so that the continuum limit is accurate if the DM interaction is
much weaker than the Heisenberg interaction.

The presence of disorder, magnetic vacants, etc. will merely have the effect of changing the value of the
parameters entering the free energy~(\ref{eq:f0}), 
as far as the condition which guarantees the validity of the continuum limit is
not violated. Hence, the continuum model can describe different samples of the same material by tuning
properly the parameters entering $f_0$, which can be used to fit the experimental results corresponding 
to each individual sample. In particular, $\mu^2$ has to be large to ensure the validity of the
continuum limit, but otherwise can be adjusted to reproduce the experimental phase diagram.
Reproducing the experimental results directly from the full hamiltonian given by
Eqs.~(\ref{eq:hamilFM})-(\ref{eq:hamilZeeman}), without any simplification, 
is extremely difficult.

We only deal with magnetic fields perpendicular to the DM axis, so that $H_z=0$, and without any
loss we can take $H_y=0$ and consider only $H_x$, since the continuum free energy has rotational symmetry
around the DM axis ($\hat{z}$). The strength of the single-ion anisotropy is chosen in such a way
that the relation $H_{z0}\approx 10H_{x0}$ between the parallel and perpendicular critical 
fields observed in \CrNbS\ holds, what is ensured by setting $\gamma=2.58 q_0^2$. This value is
obtained from the low T critical perpendicular and parallel fields \cite{Laliena16}, given respectively
by $\beta_{x0}=(\pi^2/16)q_0^2$ and $\beta_{z0}=q_0^2+2\gamma$.

Computations performed with $\gamma=0$ showed that, as the field is 
purely perpendicular, the single-ion anisotropy does not have much influence on the results.
In the computations we set $q_0=1$, what merely amounts to a choice of the unit lenght.

The minimum of $\mathcal{F}_0$ is a solution of the corresponding Euler-Lagrange equations.
As the minimum depends only on $z$, and $M_z=0$, they read 
\begin{equation}
\vec{M}^{\prime\prime} = \Omega \vec{M}^\prime + \Phi\vec{M}
+ 2q_0\hat{z}\times[\vec{M}^\prime-(\Omega/2)\vec{M}]- (\beta_x/F)\hat{x},
\label{eq:EL}
\end{equation}
\noindent
where the prime stands for the derivative with respect to $z$ and, with $F_k=d^kF/dM^k$, $G=F+MF_1$,
and $M^\prime=dM/dz$, we have $\Omega=-2(F_1/F)M^\prime$ and
\begin{widetext}
\begin{equation}
\Phi = (F_1/MG)\left[M^{\prime2} -\vec{M}^{\prime2}
+(2F_1/F-F_2/F_1)MM^{\prime2}
+2q_0\hat{z}\cdot(\vec{M}\times\vec{M}^\prime)
+\gamma/G
+\vec{\beta}\cdot\vec{M}/F\right] + (\alpha-\mu^2q_0^2F)/G.
\label{eq:Phi}
\end{equation}
\end{widetext}

\begin{figure}[t!]
\centering
\includegraphics[width=\linewidth,angle=0]{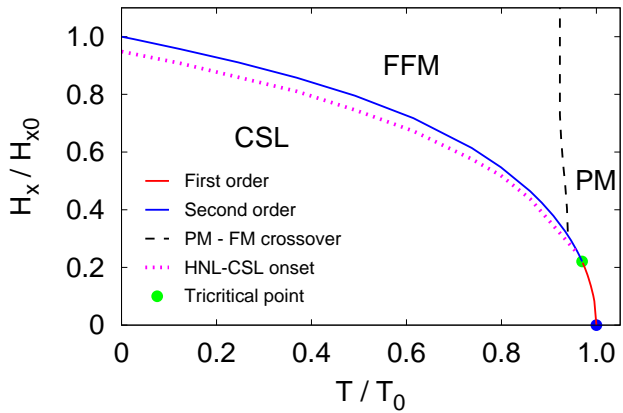}
\caption{H-T phase diagram calculated for $\mu^2=120$ and $\gamma=2.58 q_0^2$. 
The blue and red lines correspond, respectively with a second and first order phase transition.
The pink dotted line marks the onset of the highly non-linear CSL.
The transition lines described with the dimensionless variables $T/T_0$ and $H_x/H_{x0}$
are almost insensitive to $\mu^2$ if this parameter is large. 
The position of the tricritical point, however, varies appreciably $\mu^2$ 
(see Fig.~\ref{fig:tricrit_mu2}).
The zero-temperature critical field $H_{x0}$ also depends on $\mu^2$. 
\label{fig:phd}}%
\end{figure}

\begin{figure}[t!]
\centering
\includegraphics[width=\linewidth,angle=0]{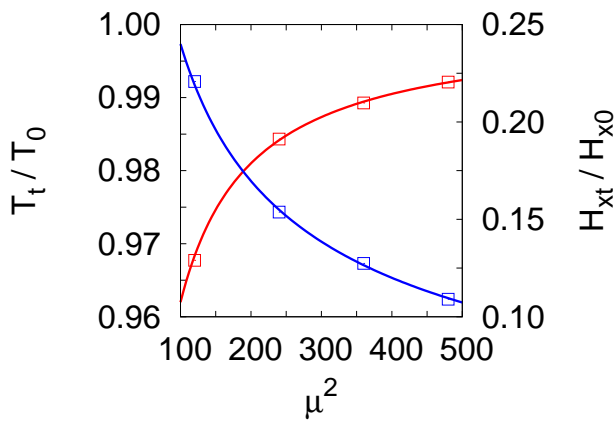}
\caption{The position of the tricritical point as a function of $\mu^2$ for 
$\gamma=2.58 q_0^2$. The open 
squares are computed points and the solid lines the result of fits.  
$T_t/T_0$ is given by the red squares and corresponds to the left ordinate scale.
The red solid line is $T_t/T_0=1-3.8/\mu^2$.
$H_{xt}/H_{x0}$ is given by the blue squares and corresponds to the right ordinate scale.
The blue solid line is $H_{xt}/H_{x0}=2.4/\sqrt{\mu^2}$.
\label{fig:tricrit_mu2}}%
\end{figure}

The general solution of the system of two second order differential equations for $M_x$ and $M_y$
(\ref{eq:EL}) contains four arbitrary integration constants. The task is to find the particular solution 
which minimises $\mathcal{F}_0$ following the method described in Ref.~\onlinecite{Laliena16}.
On physical grounds, we expect a periodic ground state, with period $L$. 
Hence, the free energy density $\bar{f}_0=\mathcal{F}_0/V$, where $V$ is the volume, is equal to the 
free energy averaged over one period, that is $\bar{f}_0=(1/L)\int_0^Lf_0(z)dz$, and the boundary conditions 
are $\vec{M}(0)=\vec{M}(L)$. Since the equations are second order periodicity requires also the 
equality of the first derivatives at the boundaries, \textit{i.e.} $\vec{M}^\prime(0)=\vec{M}^\prime(L)$. 
These additional conditions cannot be generally imposed on the boundary value problem, since it would 
be overdetermined. 
We used then the following strategy: with no loss, set $M_y(0)=M_y(L)=0$,
and for given $L$ and $M_x(0)$, solve numerically the boudary value problem; 
for fixed $L$, tune $M_x(0)$
until periodicity is reached; then, compute $\bar{f}_0$ via a numerical quadrature algorithm. 
The physical period $L$ is the minimum of $\bar{f}_0$.

\section{Phase diagram \label{sec:phd}}

To obtain the phase diagram we compare the free energies of the FFM (or PM) and CSL states.
The FFM state is always a solution of Eqs.~(\ref{eq:EL}) and its magnetization
obeys the equation 
\begin{equation}
\gamma F_1/G -M(\mu^2q_0^2F-\alpha) =\beta_x. 
\end{equation}
For $\beta_x=0$ the problem can be analytically solved. At low $\alpha$ the
free energy is minimized by an helix with pitch $L_0=2\pi/q_0$ independent of $\alpha$, 
and modulus $M_0$, which is the solution of $F(M_0)=\alpha/(\mu^2q_0^2)$ and
decreases monotonically with $\alpha$ from $M_0=\infty$ 
(what implies saturation of magnetization) at $\alpha=0$
to $M_0=0$ at 
\begin{equation}
\alpha_0=(\mu^2+1)q_0^2/3+2\gamma/15.
\end{equation}
Above $\alpha_0$ the ground state is PM. The transition takes place continuously and
$M_0$ vanishes as a power law: $M_0\sim(\alpha_0-\alpha)^{1/2}$. It is obviously an
instability type transition.
For $\alpha=0$ the problem has also been analytically solved\cite{Dzyal64}.
The transition from the CSL to the FFM state takes place continuously as the period
of the CSL diverges at critical field $\beta_{x0}=(\pi^2/16)q_0^2$. It is a
nucleation type transition. 

\begin{figure}[t!]
\centering
\includegraphics[width=\linewidth,angle=0]{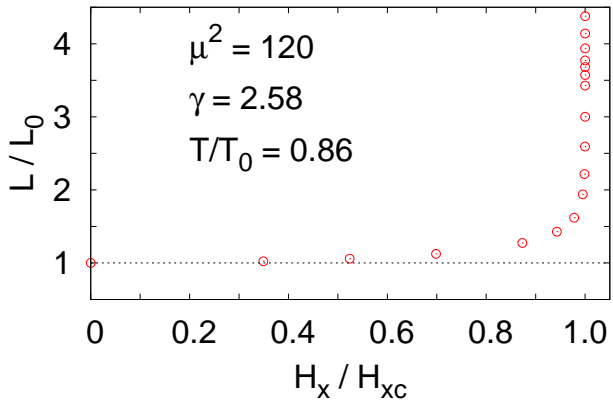}
\caption{The period of the CSL as a function of the magnetic field at 
constant temperature, below the tricritical point. Notice that it diverges at the
critical field $H_{xc}$. The period is normalized by the zero-field
period, $L_0=2\pi/q_0$.
\label{fig:L}}%
\end{figure}

The phase diagram for large $\mu^2$ is displayed in Fig.~\ref{fig:phd}.
The magnetic field is normalized by the critical field at $T=0$, 
\begin{equation}
H_{x0} = (k_BT_0/gS\mu_B)\beta_{x0}/\alpha_0, \label{eq:Hx0} 
\end{equation}
where $T_0$ is the zero-field critical temperature.
At low $T$ the CSL state continuously approaches the FFM as its 
period diverges, and the transition is continuous, of nucleation type. 
In the vicinity of $T_0$, however, the transition is discontinuous, the two states 
coexist on the transition line, and both are present in its neighborhood, one as 
stable and the other as metastable state. 
The continuous and discontinuous transition lines are separated by a 
tricritical point, $(T_t,H_{xt})$. The other end of the discontinuous transition line
is the zero-field critical point, $T_0$, where the transition is of instability type.
The appearance of first order transitions and tricritical points is thus related to
the change from nucleation to instability type continuous transitions.
A similar behaviour has been obtained in the zero temperature phase diagram 
with oblique magnetic field \cite{Laliena16}.

We shall denote by $H_{xc}(T)$ or by $T_c(H_x)$ the transition field or temperature at 
given $T$ or $H_x$, respectively. In terms of the dimensionless magnitudes 
$T/T_0 = \alpha/\alpha_0$ and $H_x/H_{x0}=\beta_x/\beta_{x0}$ 
the shape of the transition line is nearly independent of $\mu^2$ for large $\mu^2$.
However, the position of the tricritical point $(T_t/T_0,H_{xt}/H_{x0})$
on the line does depend appreciably on $\mu^2$.
Fig.~\ref{fig:tricrit_mu2} displays $T_t/T_0$ and $H_{xt}/H_{x0}$ as a function of
$\mu^2$. A fit of the computed points (open squares) shows that to high accuracy 
the position of the tricritical point is given by the equations 
$T_t/T_0=1-3.8/\mu^2$, $H_{xt}/H_{x0}=2.4/\sqrt{\mu^2}$.  These functions are represented by the 
continuous lines in Fig.~\ref{fig:tricrit_mu2}.
The complete information about the phase boundary for large $\mu^2$ is thus contained
in Figs.~\ref{fig:phd} and~\ref{fig:tricrit_mu2}.
Unless stated otherwise, the results shown from now on are obtained for $\mu^2=120$.

Two tricritical points appear in the magnetic phase diagram defined by the
perpendicular and parallel magnetic field components $(H_x,H_z)$ at zero 
temperature~\cite{Laliena16}. The one labeled as TC2 in Ref.~\onlinecite{Laliena16} 
has the same features as the tricritical point found here, as it separates a line
of continuous transtions of nucleation type from a line of discontinuous transitions.   
A tricritical line connecting these two points is then expected in the
tridimensional phase diagram $(T,H_x,H_z)$. Although the method of this
work is not valid at very low $T$, the qualitative features of the phase diagram will probably
remain valid as they interpolate from the zero $T$ limit to the high $T$ regime.

\section{Structure of the chiral Soliton Lattice \label{sec:csl}}

At zero field and $T<T_0$ the ground state is an helix with a period $L_0=2\pi/q_0$.
The effect of the field is to deform the helix to a CSL and to increase the period, $L$, 
which, at non-zero field, increases also with temperature.
Initially the growth of $L$ is very modest. However, beyond a certain field, or temperature, 
it increases rapidly. The crossover takes place roughly in the region where 
the curve defined by the period as a function of field (or temperature) has maximum curvature. 
Fig.~\ref{fig:L} illustrates the behaviour of $L$ for $T/T_0=0.86$. 
In this case the crossover takes place around $H_x/H_{xc}\approx 0.95$.

\begin{figure}[t!]
\centering
\includegraphics[width=\linewidth,angle=0]{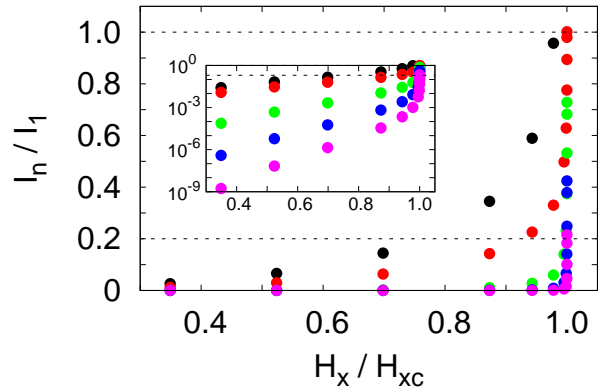}%
\caption{The intensity of the zero-mode (black circles) and the second (red), third (green),
fourth (blue), and fifth (pink) harmonics of the local magnetization ($\vec{m}$)
normalized by the
intensity of the first harmonic (fundamental wave-vector), 
as a function of the magnetic field for fixed temperature
($T/T_0=0.86$). The intensity of the third and higher order harmonics starts to be 
appreciable when
the intensity of the second harmonic reaches roughly 20\% (lowe dashed line) 
of the intensity of the first harmonic. 
Notice that the intensity of all harmonics becomes equal as the
transition point is approached. The inset displays the same data in logarithmic scale.
\label{fig:harmonics}}%
\end{figure}

The presence of two regimes suggested by the behaviour of the period and other quantities 
as magnetization (see section \ref{sec:sing}), can be understood by an analysis of the 
spatial variation of the local magnetic moment, $\vec{m}(z)$. Let us expand it in 
Fourier modes,
$\vec{m}(z) = \sum_n \vec{m}_n \exp(\mathrm{i} n 2\pi z/L)$, and define the
intensity of the n-th harmonics as $I_n=|\vec{m}_n|^2$. The behaviour of the intensity
of the first five fourier modes, including the zero-mode, $n=0$, normalized by the 
first harmonic intensity, is displayed
in Fig.~\ref{fig:harmonics} as a function of the field for $T/T_0=0.86$. 
The relative intensities of the higher order harmonics are very small 
for $H_x/H_{xc}<0.5$ and the CSL is actually a slightly distorted helix. 
In the region from $H_x/H_{xc}\approx 0.5$ to $H_x/H_{xc}\approx 0.95$ the first and the 
second harmonics give the main contribution, and the intensities of the higher harmonics 
are negligible. Structures of this kind, characterized essentially by only 
one or two harmonic modes plus the zero mode may be 
called a quasilinear CSL. Finally, for $H_x/H_{xc}>0.95$ the intensity of the second harmonic 
grows very rapidly and higher harmonics also develop rapidly: a \textit{highly nonlinear} 
CSL (HNL-CSL) appears. The onset of highly nonlinearity coincides approximately with the 
point of maximum curvature of the curve defined by the intensity of the second harmonic 
versus the field. Notice that it is roughly the point where the intensity of the second 
harmonic reaches 20\% of the first harmonic intensity, and it also coincides with the
point of change of regime in the behaviour of the period discussed above.

The zero mode remains finite while all the higher harmonics, including the
first one, tend to zero as the transition point is approached, and $I_0/I_1$ diverges.
That is why on the scale of Fig.~\ref{fig:harmonics} the zero-mode is not seen in the
vicinity of the transition point.

\begin{figure}[t!]
\centering
\begin{minipage}{0.25\textwidth}
\centering
\includegraphics[width=\linewidth,height=\linewidth,angle=0]{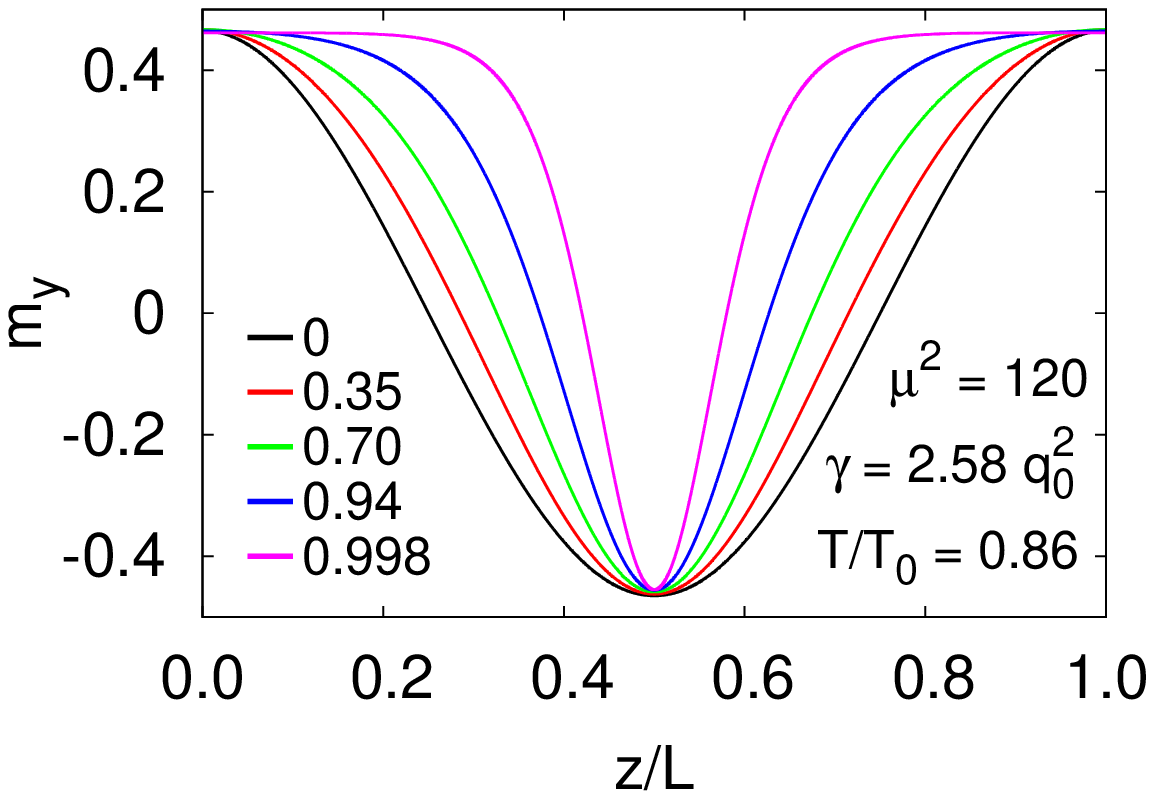}%
\end{minipage}%
\begin{minipage}{0.25\textwidth}
\centering
\includegraphics[width=\linewidth,height=\linewidth,angle=0]{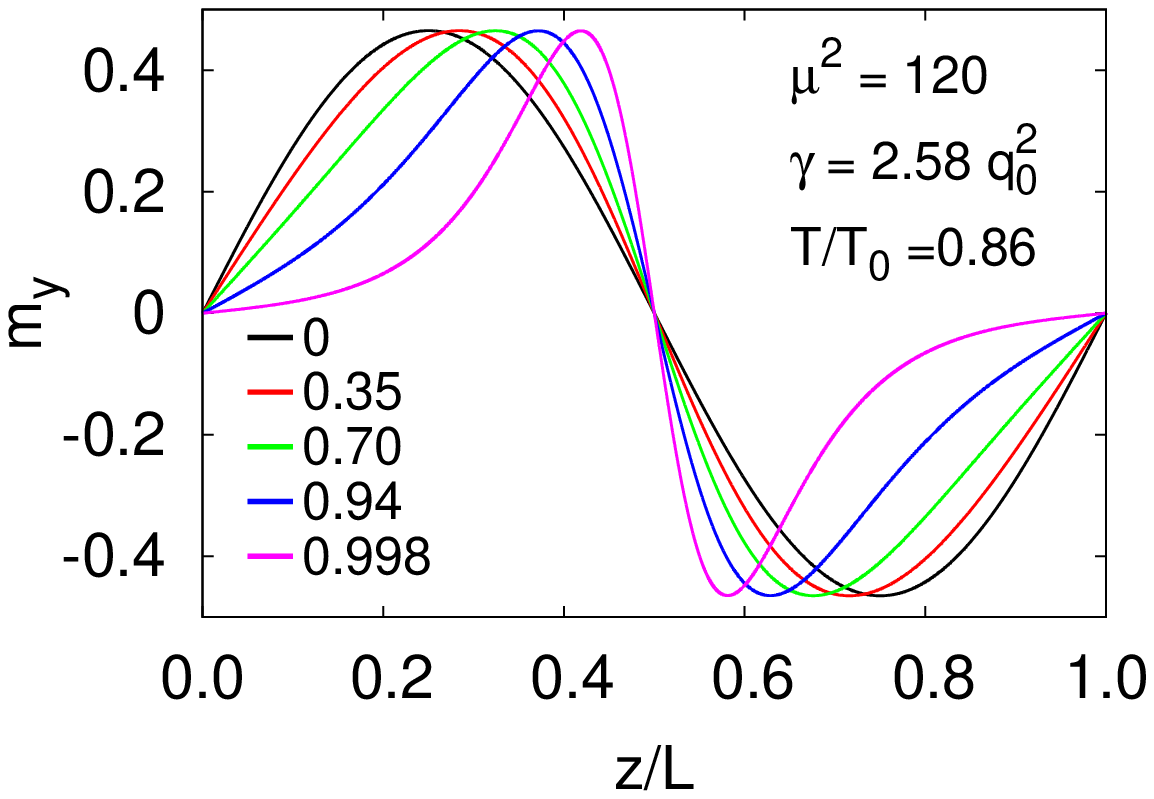}%
\end{minipage}
\caption{
The spatial variation of the components of the local magnetization $m_x$ (left) and $m_y$ 
(right) for fixed temperature and several values of $H_x/H_{xc}$
displayed in the legend, which correspond to different typical regions of the phase diagram
(see text, section~\ref{sec:csl}).
\label{fig:m}}%
\end{figure}

Incidentally, notice that these results imply that, for $T/T_0=0.86$, perturbative 
calculations keeping only the zero-mode and the first harmonic will be accurate 
for $H_x/H_{xc}<0.5$; if the second harmonic is also taken into account, the perturbative 
computation will be accurate for $H_x/H_{xc}<0.95$. Above $H_x/H_{xc}\approx 0.95$, the full 
nonperturbative computation is necessary.

It is also illustrative to visualize the spatial variation of $\vec{m}(z)$ in some 
typical cases, displayed in Fig.~\ref{fig:m}. The black lines correspond to the helix at zero 
field; the red lines ($H_x/H_{xc}=0.35$), to an slightly distorted helix; the the green lines
($H_x/H_{xc}=0.7$) to a quasilinear CSL with only two non negligible harmonics; 
the blue lines ($H_x/H_{xc}=0.94$) roughly to the HNL-CSL onset;
and the pink lines ($H_x/H_{xc}=0.998$) to a HNL-CSL state. The spatial variation of the 
modulus of $\vec{m}$, displayed in Fig.~\ref{fig:edens} (left),
is small, but its importance increases as the transition point is approached.
The right panel of Fig.~\ref{fig:edens} shows the spatial variation of the free energy density.

The HNL-CSL onset defined above is signaled by the pink dotted line in 
Fig.~\ref{fig:phd}. It could be related to some experimentally detected anomalies 
of the magnetization and the AC susceptibility reported in Refs.~\onlinecite{Ghimire13}  
and~\onlinecite{Tsuruta16}, respectively. Notice that the HNL-CSL develops only for fields 
above the tricritical field and temperatures below the tricritical temperature, 
in good agreement with the experimental results of Ref.~\onlinecite{Tsuruta16}.

\section{Soliton density}

\begin{figure}[t!]
\centering
\begin{minipage}{0.25\textwidth}
\centering
\includegraphics[width=\linewidth,height=\linewidth,angle=0]{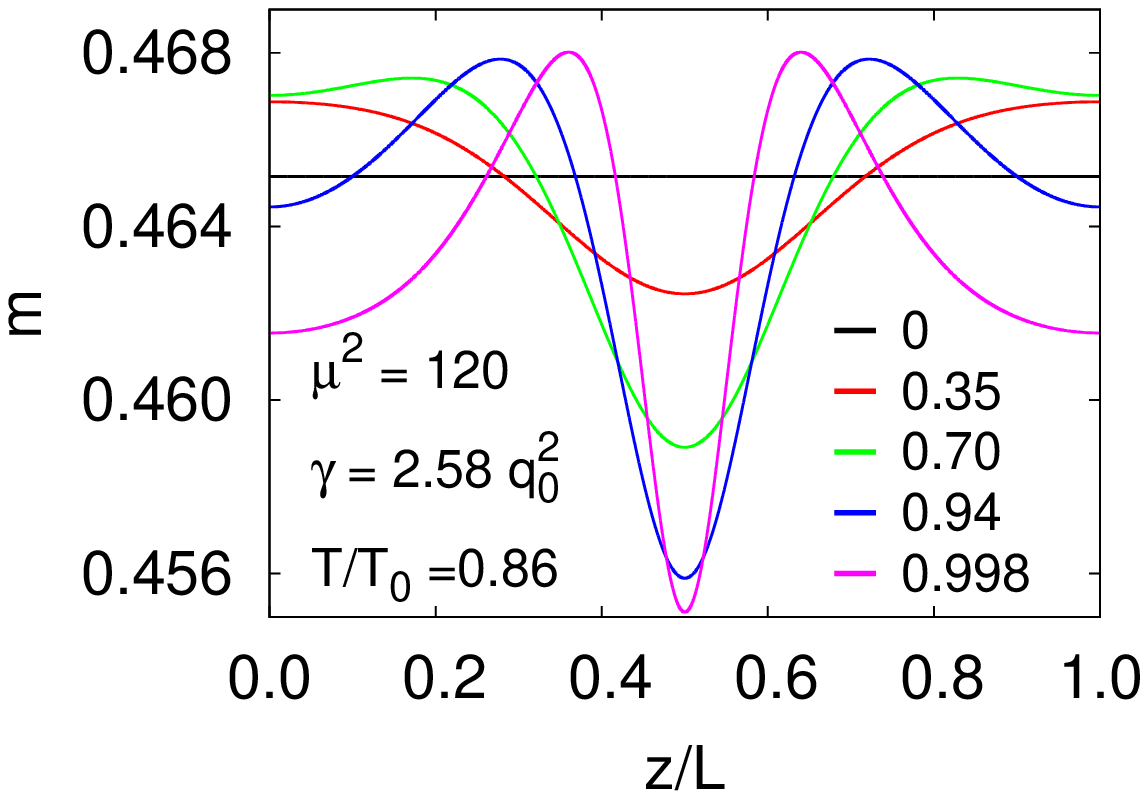}%
\end{minipage}%
\begin{minipage}{0.25\textwidth}
\centering
\includegraphics[width=\linewidth,,height=\linewidth,angle=0]{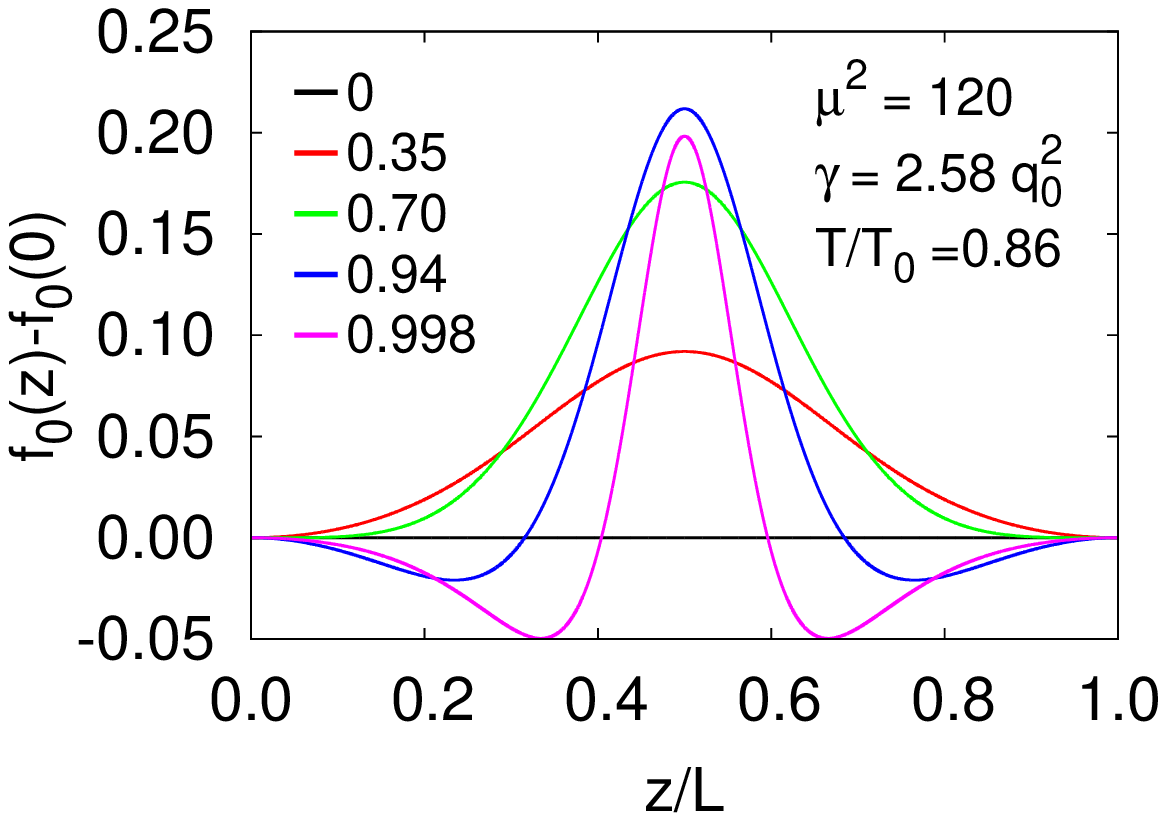}%
\end{minipage}
\caption{The spatial variation of the modulus of the local magnetic moment $m$ (left)
and the free energy density $f_0$ (right) for fixed temperature and several values of $H_x/H_{xc}$
displayed in the legend, which correspond to different typical regions of the phase diagram
(see text, section~\ref{sec:csl}).
\label{fig:edens}}%
\end{figure}

\begin{figure}[t!]
\centering
\includegraphics[width=\linewidth,angle=0]{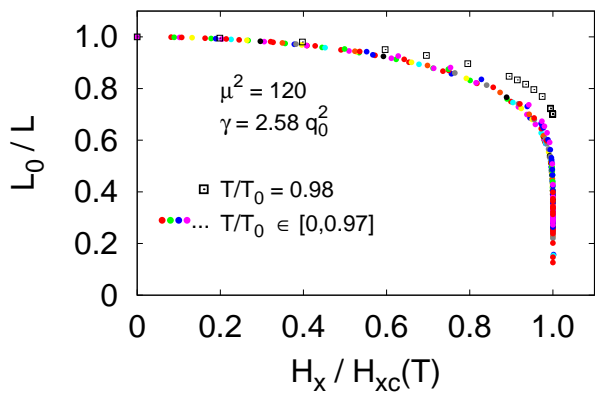}%
\caption{Soliton density ($L_0/L$) versus normalized field for 14 values of $T$ below the 
tricritical point (colored circles) and one between the tricritical temperature and $T_0$, 
where the transition is discontinuous (open black squares). Observe that the soliton density 
curves are universal below the tricritical temperature, but not above.
\label{fig:univ}}%
\end{figure}

For temperatures below the tricritical point the soliton density, $L_0/L$, is a nearly universal 
function, independent of $T$, of the dimensionless reduced
field $H_x/H_{xc}$, as can be seen in Fig.~\ref{fig:univ}. This fact explains the universality of 
the magnetoresistance curves of Ref.~\onlinecite{Togawa13}. The universality of the
soliton curves below the tricritical temperature 
can be understood in the light of the modest spatial modulation
of the local magnetic moment modulus, $m$ (Fig.~\ref{fig:edens}, left). 
Although in this work the full modulation of $m$
has been taken into account, an approximate calculation ignoring the $z$ dependence of
$m$ is a good approximation except in the close vicinity of the phase transition line.
This approximation leads to a sine-Gordon equation similar to the zero temperature 
case \cite{Dzyal64}, with an effective magnetic field given by $\beta_x/m(\alpha)$.
Within this approximation, which will be discussed further in section~\ref{sec:disc},
the soliton density curves at any temperature are given by a unique function of $H_x/H_{xc}$.

Above the tricritical temperature the universality of the soliton density
curves is lost. This is illustrated in Fig.~\ref{fig:univ} by the black open squares, which 
correspond to a temperature $T/T_0=0.98$, higher than the tricritical temperature. 
In these cases the transition to the FFM state takes place discontinuously 
from a quasilinear CSL, before the HNL-CSL is formed. 
The approximation that ignores the spatial modulation of $m$ fails qualitatively on this part 
of the phase diagram: it locates the transition point line rather accurately but it predicts
a second order nucleation type transition and universality of the soliton density curves. 
The lack of universality is thus a signal of the first order transition and can be used experimentally 
to locate the tricritical point. Strictly speaking, it may also be the signal a second order instability
type transition. In any case, if this were the case, a singular point separating the 
nucleation and instability transition lines would appear in the phase diagram.

The soliton density $L_0/L$ is an order parameter for the transition. Its behaviour along the 
transition line, parametrized by $T/T_0$, is displayed in Fig.~\ref{fig:latHeat} (left). 
Along the first order line the soliton density drops discontinuously to zero from a finite 
value. The gap vanishes at the tricritical point with a power law singularity, with 
exponent 1/4 (Fig.~\ref{fig:latHeat} left).

\section{Singularities along the transition line \label{sec:sing}}

\begin{figure}[t!]
\centering
\begin{minipage}{0.25\textwidth}
\centering
\includegraphics[width=\linewidth,height=\linewidth,angle=0]{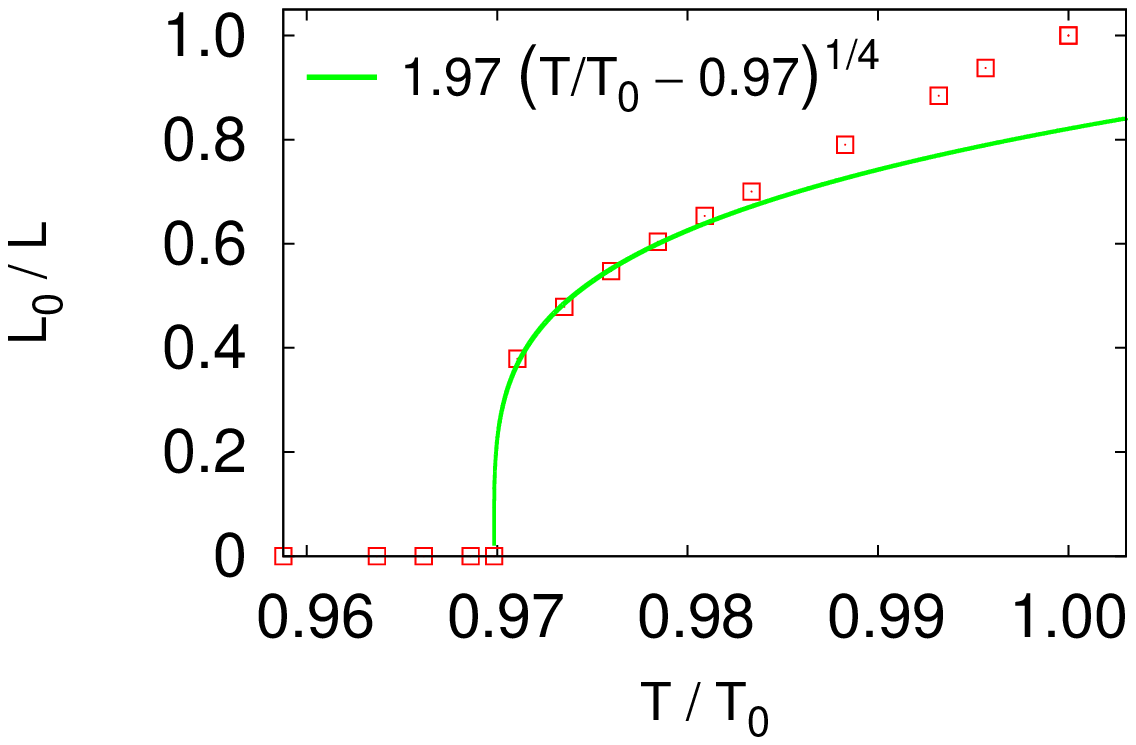}%
\end{minipage}%
\begin{minipage}{0.25\textwidth}
\centering
\includegraphics[width=\linewidth,height=\linewidth,angle=0]{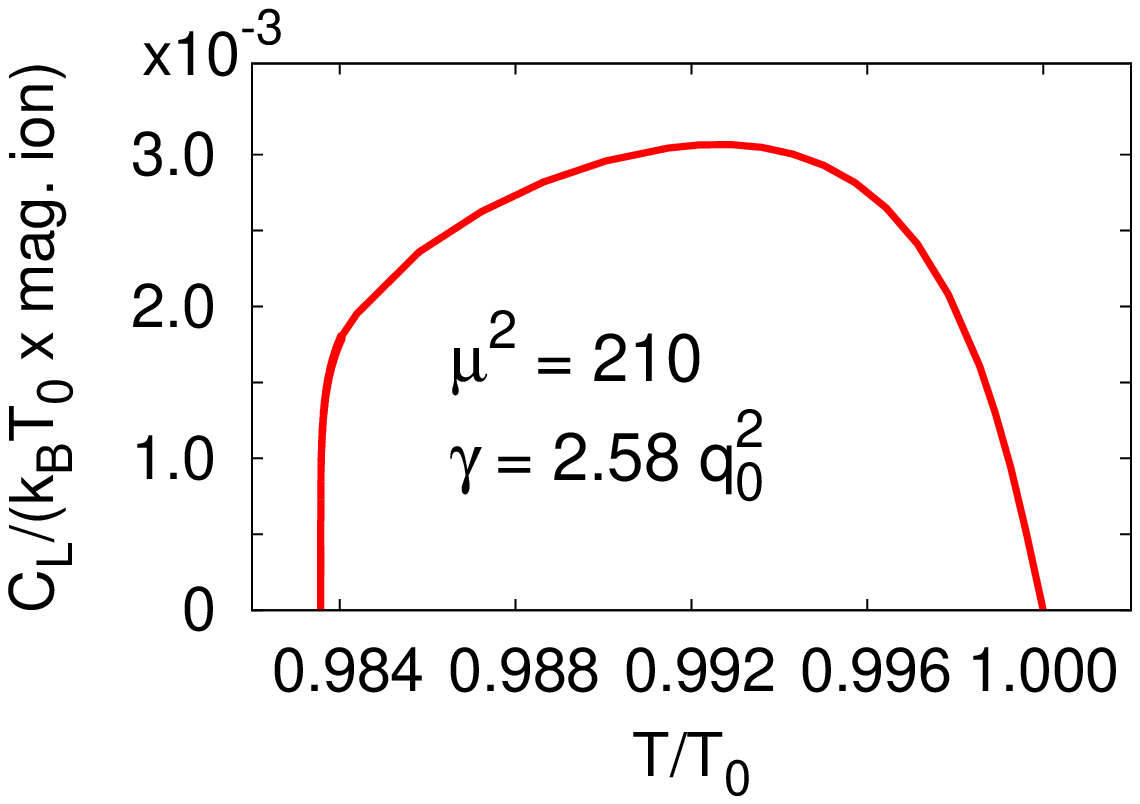}
\end{minipage}
\caption{
Left: soliton density discontinuity along the transition line; the results suggest that
the gap vanishes at the tricritical point as a power law with exponent 1/4.
Right: the latent heat along the first order transition line.
\label{fig:latHeat}}%
\end{figure}

The latent heat along the first order line vanishes at its two end points, the zero field
critical point, $T_0$, and the tricritical point, $T_t$, and therefore it reaches a maximum at some point
on the line. Its behaviour for $\mu^2=210$ and $\gamma=2.58q_0^2$ is shown in 
Fig.~\ref{fig:latHeat}.

\begin{figure}[t!]
\centering
\includegraphics[width=\linewidth,height=0.4\linewidth,angle=0]{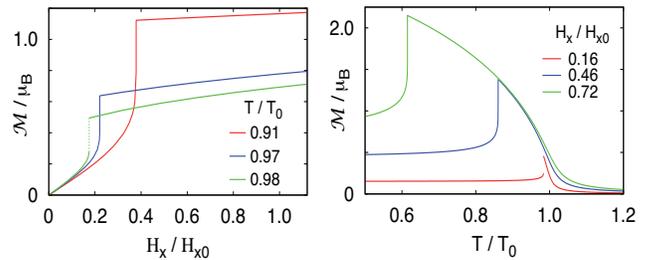}%
\caption{
Magnetization per magnetic ion as a function of $H_x$ for fixed $T$ (left) and 
as a function of $T$ for fixed $H_x$ (right).
\label{fig:magnet}}%
\end{figure}

The behaviour of the magnetization, 
\begin{equation}
\mathcal{M}=g\mu_BS\left|\frac{1}{L}\int_0^L\vec{m}(z) dz\right|,
\end{equation}
as a funtion of the magnetic field for fixed temperature and as a function of temperature for fixed 
field is displayed in Fig.~\ref{fig:magnet}. The unusual increase of the magnetization with
temperature is a distinct feature of the CSL state \cite{Kishine15} which can be understood as follows.
As $T$ increases the modulus of the local magnetic moment, $m$, decreases.
The order of magnitude of the DM energy is proportional to $m^2$ while the Zeeman energy is
proportional to $m$. Therefore, as $T$ increses the relative importance
of the Zeeman energy with respect to the DM energy increases
and the spins tend to be more aligned with the field, increasing
the period of the CSL and the magnetization. 
This effect overcomes the decrease of magnetization
due to the decrease of $m$, and the net effect is the growth of the magnetization with $T$.

\begin{figure}[t]
\centering
\begin{minipage}{0.25\textwidth}
\centering
\includegraphics[width=\linewidth,height=\linewidth,angle=0]{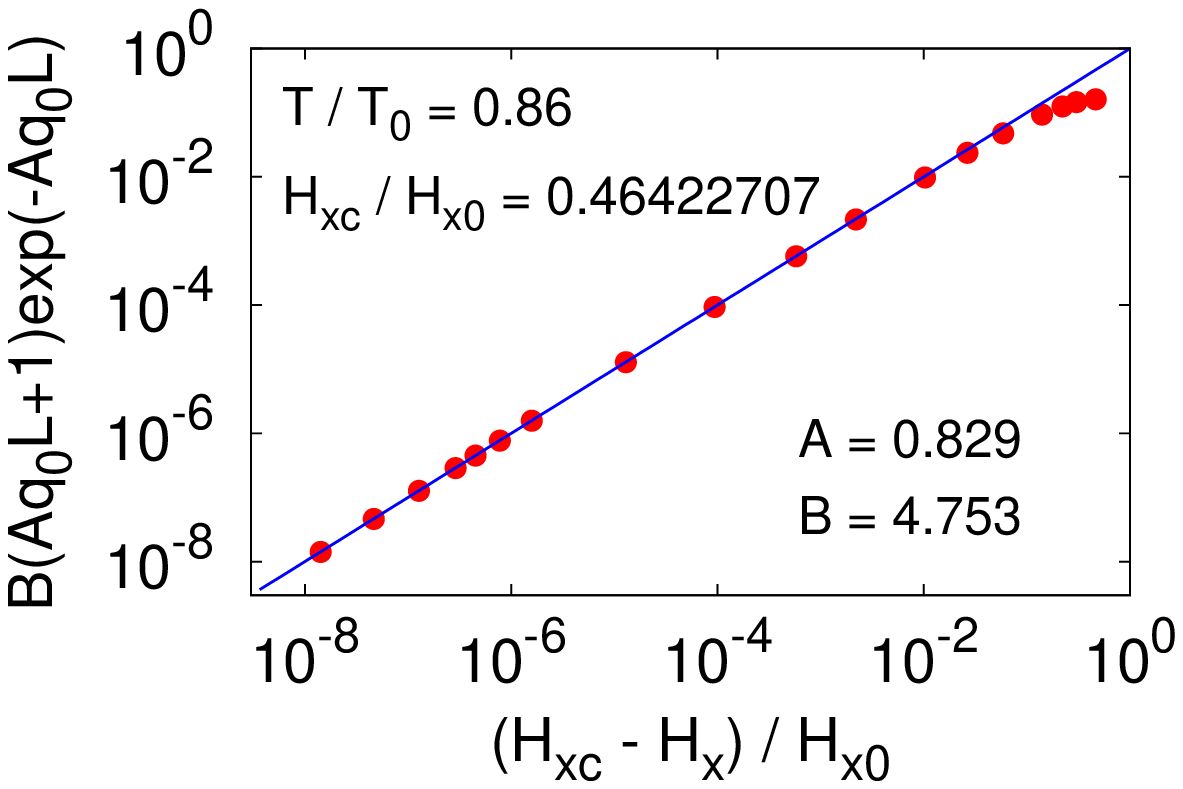}%
\end{minipage}%
\begin{minipage}{0.25\textwidth}
\centering
\includegraphics[width=\linewidth,,height=\linewidth,angle=0]{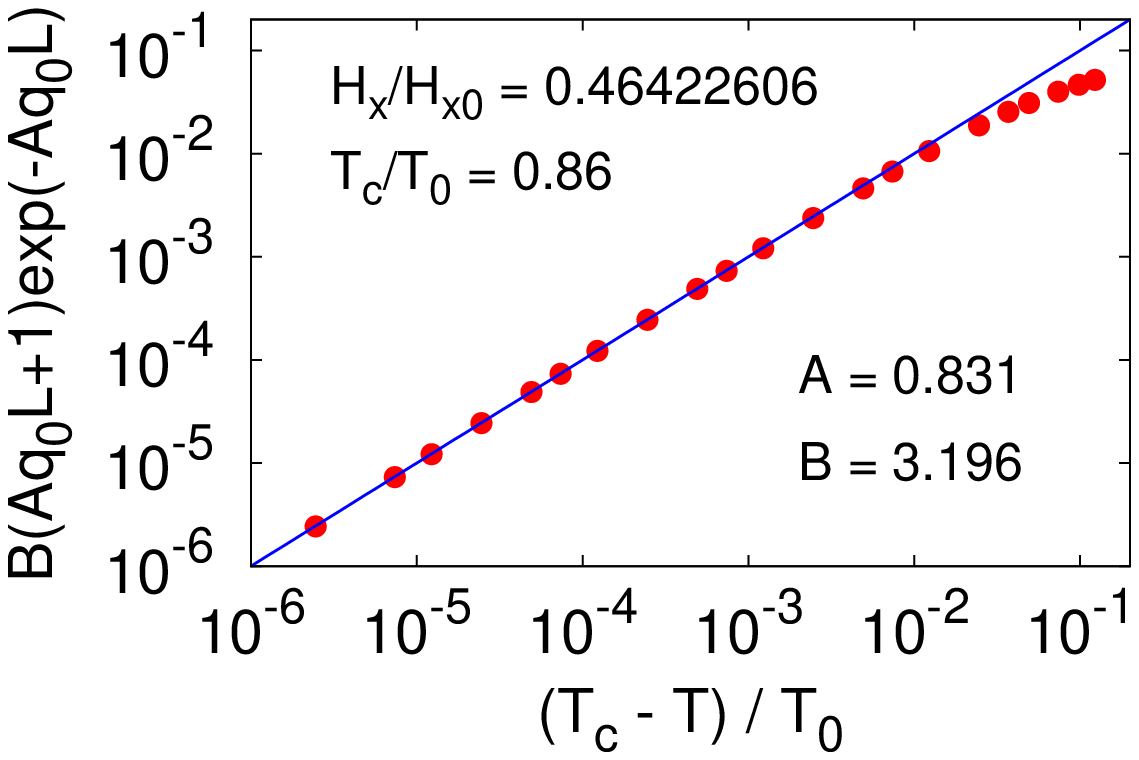}%
\end{minipage}
\caption{Scaling of the period as the continuous transition line is approached 
along the constant $T/T_0=0.86$ line (left) and the constant $H_x/H_{x0}=0.46422606$ line (right).
The blue lines represent the $y=x$ straight line.
\label{fig:scal}}%
\end{figure}

\begin{figure}[t!]
\centering
\begin{minipage}{0.25\textwidth}
\centering
\includegraphics[width=\linewidth,,height=\linewidth,angle=0]{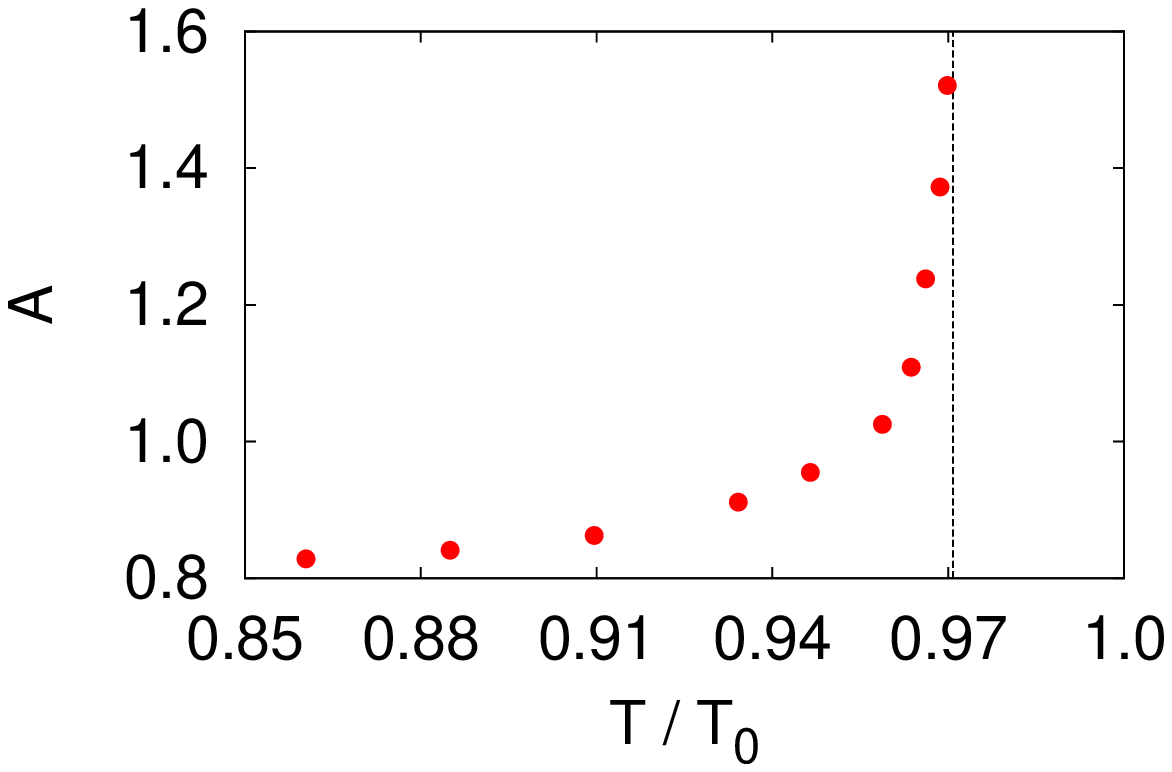}%
\end{minipage}%
\begin{minipage}{0.25\textwidth}
\centering
\includegraphics[width=\linewidth,height=\linewidth,angle=0]{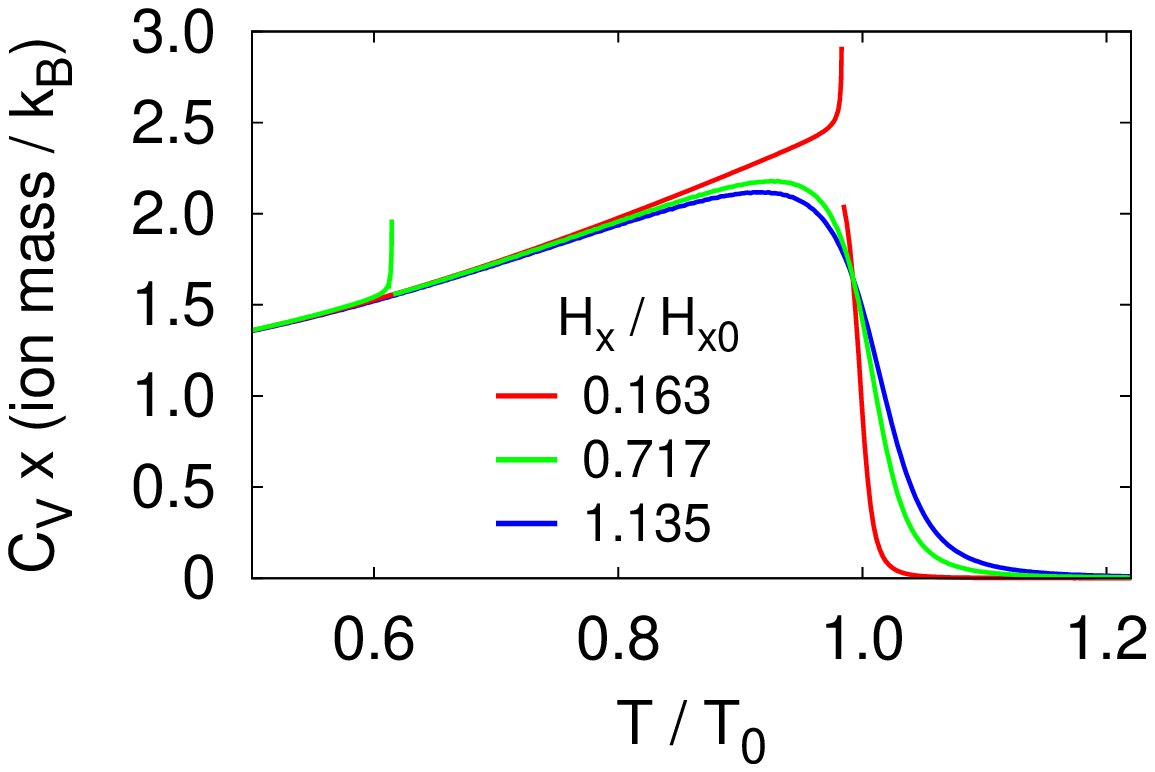}%
\end{minipage}
\caption{
Left: coefficient $A$ of the scaling law of the period;
it diverges at the tricritical point.
Right: specific heat as a f<unction of temperature at fixed field in three cases: 
first (red) and second (green) order transitions, and no transition (blue).
The broad shoulder signals the crossover from PM to FM behaviour; 
the narrow peak to the CSL-FFM phase transition. 
\label{fig:sh}}%
\end{figure}

The magnetization shows a finite jump on the first order line, while it is continuous on the 
second order line (Fig.~\ref{fig:magnet}), where it presents a singularity 
which is controlled by the divergence of $L$, since the difference between 
the magnetization on the CSL and FFM phases scales as $1/L$.
The numerical results show that when the transition point is approached 
keeping $T$ constant $L$ satisfies the scaling law 
\begin{equation}
B(Aq_0L+1)\exp(-Aq_0L)\sim (H_{xc}-H_x)/H_{x0}.
\end{equation}
This scaling law, which also holds along the continuous transition line in 
the $(H_x,H_z)$ plane at $T=0$ \cite{Laliena16}, is motivated by the scaling law at $T=0$, 
given by 
\begin{equation}
(\sqrt{\beta_{x0}}L+1)\exp(-\sqrt{\beta_{x0}}L)\sim (\beta_{x0}-\beta_x)/8\beta_{x0}.
\end{equation}
Fig.~\ref{fig:scal} (left) displays the scaling of $L$ for $T/T_0=0.86$. 
The same scaling law holds if the transition line is approached by keeping 
$H_x$ constant, with the right hand side scaling variable substituted by 
$(T_c-T)/T_0$.
Fig.~\ref{fig:scal} (right) displays this scaling for $H_x/H_{x0}=0.46422606$, 
which corresponds to the transition at $T/T_0=0.86$. 
Therefore the scaling of $L$ is a universal feature of the CSL.
The coefficient $A$ is independent of the direction (constant $T$ or $H_x$) along which the 
transition point is approached, and it is thus a feature of the transition point; $B$, however, 
depends on the approaching direction. Both $A$ and $B$ change continuously along the 
second order line, increasing with temperature, and diverge as the tricritical point is 
approached (Fig.~\ref{fig:sh} left). 
Thus, the behaviour of $A$ and $B$ may be used to locate the tricritical point 
experimentally. 
The divergence of $A$ and $B$ means that the singular behaviour at the tricritical point is 
different from that along the continuous line, as expected. Unfortunately, it is not possible 
to determine numerically this singular behaviour without further insight. 

The specific heat diverges on the continuous transition line.
The divergence is seen as a narrow peak also observed 
by Shinozaki and collaborators~\footnote{M. Shinozaki \textit{et al.}, poster at the ChiMag2016 
Conference, Hiroshima, February 2016, 
and Y. Kato, oral presentation at the same conference.}. 
On the first order line, however, the narrow peak corresponds to a finite jump (Fig.~\ref{fig:sh}). 
The broad shoulder at higher temperature is associated to the 
crossover from PM to FFM behaviour, which is signaled by the dashed line in Fig.~\ref{fig:phd}. 
The crossover temperature is close $T_0$ for all values of the field, in agreement
with the experimental results reported in 
Refs.~\onlinecite{Ghimire13} and \onlinecite{Tsuruta16}.

\section{Relation to phenomenology \label{sec:app}}

A consequence of the universality of the continuum 
free energy is that different samples of the same material are described with a different set 
of the free energy parameters. For each sample, the parameters can be fixed from the transition
point at two temperatures: the zero-field critical temperature, $T_0$, and some lower 
temperature $T_1$ at which the phase transition is well determined. 
Let the critical field at this temperature be $H_{xc1}$. 
The phase transition line (Fig.~\ref{fig:phd}) is described by an equation of the form
$H_{xc}/H_{x0}=w(T/T_0)$, where the function $w$ is independent of $\mu^2$. Then, we have
\begin{equation}
\frac{\mu_{\mathrm{B}}H_{xc1}}{k_{\mathrm{B}}T_1} = \frac{w(T_1/T_0)}{T_1/T_0}
\frac{\mu_{\mathrm{B}}H_{x0}}{k_{\mathrm{B}}T_0}
\end{equation}
Using the the expression for $H_{x0}$ given by Eq.~(\ref{eq:Hx0}),
the following equation for $\mu^2$ is obtained:
\begin{equation}
\mu^2 = \frac{3}{gS}\frac{k_{\mathrm{B}}T_1}{\mu_{\mathrm{B}}H_{x1}} \frac{w(T_1/T_0)}{T_1/T_0}
\frac{\pi^2}{16}
-\frac{2}{5}\frac{\gamma}{q_0^2}-1.
\end{equation}
Except for $\gamma/q_0^2$, the right hand side of the above equation is completely determined 
by the experimental data $T_0$, $T_1$, and $H_{xc1}$. But $\gamma/q_0^2$ plays a minor role and 
for all samples we may use the value 2.58 obtained at low temperature \cite{Laliena16}.
A more accurate value can be obtained from measurements with
a parallel field $H_z$ (for instance, from the ratio of parallel and perpendicular critical fields). 
From the value of $\mu^2$ the position of the tricritical point is determined through the
fits displayed on Fig.~\ref{fig:tricrit_mu2}, and we get
\begin{eqnarray}
T_t &=& (1-3.8/\mu^2) T_0, \\
H_{xt} &=& \frac{2.4}{\sqrt{\mu^2}} \frac{3}{gS} 
\frac{\pi^2/16}{\mu^2+1+(2/5)\gamma/q_0^2}
\frac{k_{\mathrm{B}}}{\mu_{\mathrm{B}}} T_0
\end{eqnarray}
Thus, different samples of the same material will be described by different values of $\mu^2$
and $\gamma/q_0^2$.

\section{Discussion \label{sec:disc}}

The transition line (Fig.~\ref{fig:phd}) can be 
reproduced in a simple way by assuming that the only effect of temperature is to decrease uniformly
the value of the modulus of the local magnetic moment, $m(z)=|\langle \vec{S}(z)\rangle|/S$, 
so that $m$ is independent of $z$.
The problem thus is reduced to the solution of a chiral sine-Gordon equation with an effective 
field $\beta_x/m(\alpha,\beta_x)$. The numerical results show that in 
the CSL phase $m$ is nearly independent of $\beta_x$. Thus, neglecting the $\beta_x$ dependence of $m$,
the approximation gives by $H_{xc}/H_{x0}=m_0(T)$, where $m_0(T)=M_0F(M_0)$ and $M_0$ is the 
zero-field mean field solution (see section \ref{sec:phd}). 
This equation describes the transition line with very high accuracy. 
However, it does not capture the nature of the transitions nor the tricritical behaviour.
The local magnetic moment $m_0(T)$, obtained in the mean field approximation, may be not
accurate. However, if we consider its exact value, $|\langle\vec{S}\rangle_T|$, 
which can be measured by neutron scattering,
the approximation predicts a relationship between the critical fields and local magnetic 
moments at two different temperatures:
\begin{equation}
H_{xc}(T)/H_{xc}(T^\prime) = |\langle\vec{S}\rangle_T|/|\langle\vec{S}\rangle_{T^\prime}|.
\end{equation}

The results about the phase diagram of the monoaxial helimagnet presented here are compatible with
Monte Carlo simulations recently performed \cite{Nishikawa16}. These simulations point out that
the zero field transition is of second order and belongs to the universality class of the XY model,
and is thus of instability type,
while a different kind of transition, which might be a nucleation type second order transition,
takes place when the perpendicular magnetic field is strong enough.

Also recently Shinozaki \textit{et al.} \cite{Shinozaki15} addressed the problem of analyzing
theoretically the phase diagram of the monoaxial helimagnet with a mean field technique combined
with Monte Carlo simulations. They used a discrete model with FM exchange and DM interactions
restricted to first neighbours and a perpendicular external magnetic field. It is difficult
to compare their results with ours, as both are presented in different ways. 
They also briefly reported some signal of first order transition in the vicinity of $T_0$ but
do not locate any tricritical point.

The recent experimental findings on the phase diagram of \CrNbS\ can be understood in the 
light of the present theoretical study. For this material we have $gS\approx 3$.
The formulas of section~\ref{sec:app} can be used to make definite predictions. 
For instance, the results of Ref.~\onlinecite{Togawa12} for a crystal with $T_0\approx 127$ K and 
$H_{xc}\approx 2300\,\mathrm{Oe}$ at 110 K are reproduced with $\mu^2\approx 210$. 
Then, the tricritical point is predicted to be at 
$T_t\approx 125$ K and $H_{xt}\approx 910$ Oe.
The first order transition in the vicinity of $T_0$, the second order 
transition at lower $T$, the presence of a tricritical point at the predicted location,
and the absence of HNL-CSL for fields below $H_{xt}$
are consistent with the phase diagram reported in Ref.~\onlinecite{Tsuruta16}.
Furthermore, the universality the soliton density curves below the tricritical temperature
found here explains the universality of the 
magnetoresistance curves of Ref.~\onlinecite{Togawa13}. 
The lost of such universality is a signal of the first order transition and can be used to
locate the tricritical point experimentally. In Ref.~\onlinecite{Togawa13} magnetorresistance 
curvess are reported up to 120 K, below the predicted tricritical point.
The phase diagram drawn in Ref.~\onlinecite{Ghimire13},
which has $T_0\approx 120$~K and $H_{xc}\approx 1300$~Oe at 110 K,
is reproduced with $\mu^2\approx 310$ and a tricritical point is predicted at
119~K and 490~Oe, in a region which has not been fully explored in Ref.~\onlinecite{Ghimire13}.

\begin{acknowledgments}
We are grateful to S. Hoshino, Y. Kato, J. Kishine, Y. Masaki, and M. Shinozaki for useful discussions.
J.C. and V.L. acknowledge the grant MAT2015-68200-C2-2-P.
This work was partially supported by the scientific JSPS Grant-in-Aid for Scientific Research (S) (No. 25220803), and the MEXT program for promoting the enhancement of research universities, and JSPS Core-to-Core Program, A. Advanced Research Networks.
\end{acknowledgments}

\appendix
\section{Some details about the continuum limit \label{sec:contLim}}

Let us sketch in this appendix how the continuum limit is taken. Let us consider first the contribution of
the FM term~(\ref{eq:hamilFM}) to the mean field free energy $\mathcal{F}_0$, which is given by
\begin{equation}
-S^2\sum_{\vec{r},\vec{r}^{\,\prime}}J_{\vec{r}^{\,\prime}} \vec{m}_{\vec{r}}\cdot\vec{m}_{\vec{r}+\vec{r}^{\,\prime}}.
\label{eq:FM}
\end{equation}
Assume that $\vec{m}_{\vec{r}}$ only varies appreciably over distances much larger than the range of 
the magnetic interaction strengths $J_{\vec{r}^{\,\prime}}$ and $D_z^\prime$. 
Then, we can expand $\vec{m}_{\vec{r}+\vec{r}^{\,\prime}}$ in Taylor series around $\vec{r}$,
keeping only terms up to second order:
\begin{equation}
\vec{m}_{\vec{r}+\vec{r}^{\,\prime}} = \vec{m}_{\vec{r}} + (\vec{r}^{\,\prime}\cdot\nabla)\vec{m}_{\vec{r}} + 
(1/2)\sum_{i,j}x_i^\prime x_j^\prime\partial_i\partial_j \vec{m}_{\vec{r}}+\ldots
\label{eq:taylor}
\end{equation}
Plugging~(\ref{eq:taylor}) into~(\ref{eq:FM}) and taking into account that the term linear in
$\vec{r}^{\,\prime}$ can be cast as a total divergence and thus disappears upon summing over 
$\vec{r}$, we get, after integrating by parts and removing again a total divergence, that the FM
contribution to $\mathcal{F}_0$ is:
\begin{equation}
-JS^2\sum_{\vec{r}}\vec{m}_{\vec{r}}^2 + (1/2)S^2\sum_{\vec{r},i,j} a^2 J_{ij} 
\partial_i\vec{m}_{\vec{r}}\cdot\partial_j\vec{m}_{\vec{r}},
\end{equation}
where $J=\sum_{\vec{r}^{\,\prime}}J_{\vec{r}^{\,\prime}}$, $a^2$ is defined in Eq.~(\ref{eq:a}), and
\begin{equation}
J_{ij} = (1/a^2)\sum_{\vec{r}^{\,\prime}} x_i^\prime x_j^\prime J_{\vec{r}^{\,\prime}}.
\end{equation}
The symmetric tensor $J_{ij}$ is diagonal in an orthogonal cartesian system, which, by symmetry,
contains the DM axis, $\hat{z}$. Hence,  $J_{zz}=J$ and, as we only consider anisotropy along the DM axis,
$J_{xx}=J_{yy}=\xi J$ where $\xi$ measures the spatial anisotropy of the Heisenberg exchange interaction
and is given by
\begin{equation}
\xi = \frac{\sum_{\vec{r}^{\,\prime}}x^{\prime 2}J_{\vec{r}^{\,\prime}}}{\sum_{\vec{r}^{\,\prime}}z^{\prime 2}J_{\vec{r}^{\,\prime}}}
= \frac{\sum_{\vec{r}^{\,\prime}}y^{\prime 2}J_{\vec{r}^{\,\prime}}}{\sum_{\vec{r}^{\,\prime}}z^{\prime 2}J_{\vec{r}^{\,\prime}}}.
\end{equation}
In the continuum limit $\sum_{\vec{r}}$ is replaced
by $(1/v)\int d^3r$, where $v$ is the elementary cell volume, so that collecting
$JS^2a^2/v$ as a global factor the free energy takes the form of Eq.~(\ref{eq:F0contLim}) and
the FM contribution to the free energy density, $f_0$, is
\begin{equation}
(1/2) [ - (2/a^2) \vec{m}_{\vec{r}}^2 
+ \sum_i \xi_i \partial_i\vec{m}_{\vec{r}}\cdot\partial_i\vec{m}_{\vec{r}} ],
\label{eq:f0FM}
\end{equation}
with $\xi_x=\xi_y=\xi$ and $\xi_z=1$.

Thus, in the continuum limit there appears a contribution to $f_0$ proportional to 
$\vec{m}_{\vec{r}}^2$, which comes from the Heisenberg exchange interaction. 
Its coefficient is $1/a^2$, as can be seen in Eq.~(\ref{eq:f0FM}). 
We find convenient to define the dimensionless parameter $\mu^2 = 2/(a^2q_0^2)$,
and the coefficient of the $\vec{m}_{\vec{r}}^2$ term reads $\mu^2q_0^2/2$.

Consider now the DM interaction. The contribution of the DM term~(\ref{eq:hamilDM}) to the mean field 
free energy $\mathcal{F}_0$ is
\begin{equation}
-S^2\sum_{\vec{r},z^\prime}D_{z^\prime} \hat{z}\cdot(\vec{m}_{\vec{r}}\times\vec{m}_{\vec{r}+z^{\prime}\hat{z}})
\label{eq:DM}
\end{equation}
Plugging~(\ref{eq:taylor}) with $\vec{r}^{\,\prime}=z^\prime\hat{z}$ into~(\ref{eq:DM}) we obtain
\begin{equation}
-S^2\sum_{\vec{r},z^\prime}D_{z^\prime}\hat{z}\cdot[\vec{m}_{\vec{r}}\times\partial_z\vec{m}_{\vec{r}}
+ (1/2)z^{\prime 2}\vec{m}_{\vec{r}}\times\partial_z^2\vec{m}_{\vec{r}}].
\end{equation}
The term quadratic in $z^\prime$ is a total divergence since 
\begin{equation}
\vec{m}_{\vec{r}}\times\partial_z^2\vec{m}_{\vec{r}}=\partial_z(\vec{m}_{\vec{r}}\times\partial_z\vec{m}_{\vec{r}}),
\end{equation}
and therefore vanishes upon summing over $\vec{r}$. Hence, the contribution of the DM interaction
to $f_0$ is 
\begin{equation}
-q_0 \hat{z}\cdot (\vec{m}_{\vec{r}}\times\partial_z\vec{m}_{\vec{r}}),
\end{equation}
with $q_0$ given by Eq.~(\ref{eq:q0}).

The single-ion anisotropy~(\ref{eq:hamilAnis}) and the Zeeeman term~(\ref{eq:hamilZeeman}) do not couple
spins on different sites and thus its contribution to the continuum limit of the free energy is straightforward.
We only have to substitute $\sum_{\vec{r}}$ by $(1/v)\int d^3r$ in $\langle\mathcal{H}_{\mathrm{A}}\rangle_0$
and $\langle\mathcal{H}_{\mathrm{Z}}\rangle_0$, and extract the global factor $JS^2a^2/v$. 
The same happens with $\langle\mathcal{H}_0\rangle_0$ and $\ln\mathcal{Z}_0$. In this way we get that
the parameters entering Eq.~(\ref{eq:f0}) for $f_0$ are
\begin{eqnarray}
\gamma &=& K/Ja^2, \label{eq:gamma} \\
\vec{\beta} &=& (g\mu_\mathrm{B}/JSa^2)\vec{H}, \label{eq:beta} \\
\alpha &=& k_\mathrm{B}T/JS^2a^2. \label{eq:alpha}
\end{eqnarray}

Let us discuss briefly the conditions for the validity of the continuum limit. In~(\ref{eq:taylor}) we
neglected terms of the form 
$x_{i_1}^\prime\ldots x_{i_n}^\prime \partial_{i_1}\ldots\partial_{i_n}\vec{m}_{\vec{r}}$ with $n>2$.
They would give a contribution to $f_0$ proportional to 
$\vec{m}_{\vec{r}}\cdot\partial_{i_1}\ldots\partial_{i_n}\vec{m}_{\vec{r}}$ with a coefficient of the form
$\sum_{\vec{r}^{\,\prime}} J_{\vec{r}^{\,\prime}}x_{i_1}^\prime\ldots x_{i_n}^\prime$, which is of order $a^n$. On the other hand,
the $n$-th derivative is of the order $q_0^n$. This can be seen as follows: $\vec{m}$ is a dimensionless quantity,
hence it has to be a function of the dimensionless variable $q_0\vec{r}$. It does not depend on the other 
dimensionless variable, $\vec{r}/a$, since no modulation exist if $q_0$ vanishes (in this case the DM interaction
vanishes). Obviously, the $n$-th derivative of a function $\vec{m}(q_0\vec{r})$ is proportional to $q_0^n$.
Hence, the neglected terms are of order $(q_0a)^n=(2/\mu^2)^{n/2}$, with $n>2$,
and thus the validity of the continuum limit requires a large value of $\mu^2$.

\bibliography{references}

\end{document}